\documentclass[pre,showpacs]{revtex4}
\usepackage{amsfonts,amssymb,amsmath,latexsym,times} 
\usepackage[dvips]{graphics}

\begin{document}

[Phys. Rev. E {\bf 77}, 027201 (2008)]

\title{Leading Pollicott-Ruelle Resonances for Chaotic Area-Preserving Maps}

\author{\bf Roberto Venegeroles \footnote{E-mail: roberto.venegeroles@ufabc.edu.br}}
\affiliation{Centro de Matem\'atica, Computa\c c\~ao e Cogni\c c\~ao, Universidade Federal do ABC, 09210-170 Santo Andr\'e, SP, Brazil}

\date{\today}

\begin{abstract}

Recent investigations in nonlinear sciences show that not only hyperbolic but also mixed dynamical systems may exhibit exponential relaxation in the chaotic regime. The relaxation rates, which lead the decay of probability distributions and correlation functions, are related to the classical evolution resolvent (Perron-Frobenius operator) pole logarithm, the so called Pollicott-Ruelle resonances. In this Brief Report, the leading Pollicott-Ruelle resonances are calculated analytically for a general class of area-preserving maps. Besides the leading resonances related to the diffusive modes of momentum dynamics (slow rate), we also calculate the leading faster rate, related to the angular correlations. The analytical results are compared to the existing results in the literature.

\end{abstract}


\maketitle

\section{Introduction}

It is well known that for systems exhibiting chaotic dynamics, precise long-time predictions of individual trajectories are impossible. It is natural, therefore, to investigate the statistical properties of these systems. In this sense, the time evolution of the probability densities of trajectories $\rho_{n}$, ruled by the Perron-Frobenius (PF) operator $U$ as $\rho_{n+1}=U\rho_{n}$, have been extensively studied \cite{Gaspard,CvitanovicBook}.

Due to Liouville's theorem, $U$ can be represented by a unitary operator in a Hilbert space. Consequently, its resolvent
\begin{equation}
R(z)=\frac{1}{z-U}=\frac{1}{z}\sum_{j=0}^{\infty}U^{j}z^{-j}
\label{resRie}
\end{equation}
is singular on the unit circle in the complex $z$-plane, and the matrix elements of $R(z)$ are discontinuous there. The sum in (\ref{resRie}) is convergent for $|z|>1$ and has an analytical extension across the cut into the first Riemann sheet, which exhibits a set of singularities known as Pollicott-Ruelle (PR) resonances \cite{Pollicott,Ruelle}. To identify the PR resonances it is necessary to analytically continue the resolvent across the continuous spectrum of $U$ from the outside to the inside of the unitary circle. These resonances characterize the irreversible behavior of chaotic dynamics \cite{Gaspard,Hasegawa2}. In particular, the nontrivial ($z\neq1$) maximal PR resonance leads the exponential decay of distribution and correlation functions \cite{Khodas,Venegeroles}.

The PR resonances have attracted considerable attention not only in classical dynamics but also in quantum systems \cite{quantum}, and some numerical and semi-analytical schemes where recently developed for calculate them. Blum and Agam proposed a variational method to locate the leading resonances \cite{Blum}. Although their results describe the aparent formation of a leading quartet for two particular map cases, verified by respective numerical diagonalization of $U$, the leading resonance calculated diverges for a set of values of $K$ in the standard map case when this approach breaks down. Florido et al. extended this variational approach in a class of numerical methods in which memory function and filter diagonalization techniques are utilized by means of interpolating exponentials \cite{Florido}. Usually, there are two standard ways to calculate the PR resonances: one is based on the numerical diagonalization of the operator $U$, for which the resonances are directly calculated from its eigenvalues \cite{Blum,Weber,Sano}, the other, through the zeros of the classical Ruelle zeta-function, that is derived from the trace of the resolvent of $U$ \cite{Gaspard,CvitanovicBook}. In the last case, there are analytical calculations of these resonances for some hyperbolic systems (for which this formalism is rigourous) such as the multibaker map \cite{Gaspard}, geodesic motion in billiards of constant negative curvature \cite{Lebouefzeta}, and hard-disk scatterers \cite{Gaspard}. On the other hand, many physically realistic systems are mixed, and analytical procedures to determine resonances for this cases are thus in demand.

The motivation of the present Brief Report is to calculate analytically the leading PR resonances for slow (diffusive) and faster modes of dynamics for the general class of two-dimensional area-preserving maps:
\begin{eqnarray} 
\begin{array}{l} I_{n+1} = I_{n}+Kf(\theta_{n})\,,\\ 
                 \theta_{n+1} = \theta_{n}+c\,\alpha(I_{n+1})\qquad \mbox{mod}\, 2\pi,
\end{array} \label{map}
\end{eqnarray}
defined on the cylinder $-\pi\leq\theta<\pi,-\infty<I<\infty$. Here $f(\theta)$ is the impulse function, $\alpha(I)=\alpha(I+2\pi r)$ is the rotation number, $c$ and $r$ are real parameters, and $K$ is the stochasticity parameter. This map is commonly called the {\it radial twist map} \cite{Lichtenberg} periodic in momentum variable $I$. The specific linear rotation number (LRN) case $c\,\alpha(I)\equiv I$ for which $f(\theta)=\sin\theta$ represents the Chirikov-Taylor standard map \cite{Chirikov}, a paradigm of Hamiltonian chaos \cite{Lichtenberg}. LRN maps are periodic because $I$ can be replaced by $I$ mod $2\pi$. On the other hand, nonperiodic rotation numbers can be considered in the limit $r\rightarrow\infty$ \cite{nonG}.

\section{Projection Operators\label{Resolan}}

An usual way to determine the leading PR resonance is to evaluate the application $U^{n}$ for large values of the time $n$ when only the highest resonance survives, as it occurs for the equilibrium statistical mechanics of lattice systems. Let us consider the analysis of the resolvent (\ref{resRie}) for which $U^{n}$ can be expressed as $\oint_{C} dz R(z)z^{n}=2\pi i U^{n}$ \cite{Hasegawa2}. The spectrum of $U$ is located on the unit circle $C$ around the origin in complex $z$-plane or inside it. Thus, the contour of integration is a circle lying just outside the unit circle. In order to evaluate $U^{n}$, a very effective method based on the projection operator techniques can be used \cite{Hasegawa2,Balescu1}. In this method, we consider two mutually orthogonal idempotent operators $P$ and $Q$: 
\begin{eqnarray}
1=P+Q,\,\,\,\, P^{2}=P,\,\,\,\, Q^{2}=Q,\,\,\,\, PQ=QP=0,
\end{eqnarray}
where $1$ represents the identity operator. These operators decompose the resolvent in the following nontrivial form:
\begin{equation}
\frac{1}{z-U}=[P+Q\,\mathcal{C}(z)P]\frac{1}{z-P\mathcal{E}(z)P}[P+P\mathcal{D}(z)Q]+Q\mathcal{P}(z)Q,
\label{decresolvent}
\end{equation}
where the operators $\mathcal{P}(z)$, $\mathcal{E}(z)$, $\mathcal{C}(z)$ and $\mathcal{D}(z)$ are the discrete time version of the Brussels formalism \cite{Bandtlow}, defined by
\begin{eqnarray}
Q\mathcal{P}(z)Q&=&Q\frac{1}{z-QUQ}Q,\label{prop}\\
P\mathcal{E}(z)P&=&PUP+PUQ\mathcal{P}(z)QUP,\label{diag}\\
Q\mathcal{C}(z)P&=&Q\mathcal{P}(z)QUP,\label{crea}\\
P\mathcal{D}(z)Q&=&PUQ\mathcal{P}(z)Q.\label{dest}
\end{eqnarray}
A recent proof of (\ref{decresolvent})-(\ref{dest}) can be found at \cite{Balescu1}.

The matrix representation of the PF operator $U$ for (\ref{map}) in Fourier space $(\theta,I)\mapsto(m,q)$ is given by \cite{Venegeroles}
\begin{equation}
\left\langle m,q|U|m',q'\right\rangle=\sum_{m'}\int dq'\sum_{l}\delta(lr^{-1}-q'+q)\mathcal{G}_{l}(r,mc)\mathcal{J}_{m-m'}(-Kq'),
\label{matrixU}
\end{equation}
where the Fourier decompositions of the $\alpha(I)$ and $f(\theta)$ functions are
\begin{eqnarray}
\mathcal{G}_{l}(r,x)&=&\frac{1}{2\pi}\int d\theta\,\exp\{-i[x\alpha(r\theta)-l\theta]\}\,,\label{Gfunction}\\
\mathcal{J}_{m}(x)&=&\frac{1}{2\pi}\int\,d\theta\,\exp\{-i[m\theta-xf(\theta)]\}\,.\label{Jfunction}
\end{eqnarray}

\section{Slow Relaxation Rate\label{SLPres}}

The leading PR resonances related to diffusive modes of the momentum variable $I$ for (\ref{map}) corresponds to the relaxation rate of $PU^{n}P\sim\exp[n\gamma(q)]$ for $n\gg1$ and $P\equiv\left|0,q\right\rangle$. The diffusion coefficient $D$ is then calculated by $D=-(1/2)\partial_{q}^{2}\gamma(q)|_{q=0}$. Applying $P$ on the two sides of (\ref{decresolvent}), the projection of the PF operator $U^{n}$ can be written as

\begin{equation}
PU^{n}P=\frac{1}{2\pi i}\oint_{C}dz\frac{z^{n}}{z-\sum_{j=0}^{\infty}z^{-j}\Psi_{j}(q)},\label{relevantevolution}\\
\end{equation}
where the memory functions $\Psi_{j}(q)$ are given by \cite{Venegeroles}:
\begin{eqnarray}
\Psi_{0}(q)&=&\mathcal{J}_{0}(-Kq)\,,\label{psi0}\\
\Psi_{1}(q)&=&\sum_{m}\mathcal{J}_{-m}(-Kq)\mathcal{J}_{m}(-Kq)\,\mathcal{G}_{0}(r,mc)\,,\label{psi1}\\
\Psi_{j\geq2}(q)&=&\sum_{\{m\}}\sum_{\{\lambda\}^{\dag}}\mathcal{J}_{-m_{1}}(-Kq)\,\mathcal{J}_{m_{j}}(-Kq)\,\mathcal{G}_{\lambda_{1}}(r,m_{1}c)\nonumber\\
&\,&\times\prod_{i=2}^{j}\mathcal{G}_{\lambda_{i}}(r,m_{i}c)\mathcal{J}_{m_{i-1}-m_{i}}\left[-K\left(q+r^{-1}\sum^{i-1}_{k=1}\lambda_{k}\right)\right]\label{psij}.
\end{eqnarray}
Hereafter, the following convention will be used: the set of wavenumbers $m$ and $\left\{m\right\}=\left\{m_{1}, \dots, m_{j}\right\}$ can only take {\it non-zero} integer values, whereas the set of wavenumbers $\left\{\lambda\right\}^{\dag}$ can take {\it all} integer values, including zero, and the superscript denotes the constraint $\sum_{i=1}^{j}\lambda_{i}=0$.

The integral (\ref{relevantevolution}) can be solved by method of residues and its poles are evaluated by the well-known Newton-Raphson method: the zeros of an equation $h_{N}(z)=0$ are calculated iteratively by $z_{n+1}=z_{n}-h_{N}(z_{n})/h'_{N}(z_{n})$, where $h_{N}(z)\equiv z-\sum_{j=0}^{N}z^{-j}\Psi_{j}(q)$ assumes the truncated form of the denominator of (\ref{relevantevolution}). First, we introduce the abbreviations $M_{q}\equiv\sum^{N}_{j=0}\Psi_{j}(q)$ and $N_{q}\equiv\sum^{N}_{j=1}j\,\Psi_{j}(q)$. Notice that, taking into account the null drag condition $\int d\theta f(\theta)=0$ \cite{Venegeroles}, we have $\Psi_{0}(q\rightarrow0)=1+\mathcal{O}(q^2)$. In the general case, we have $\Psi_{j\geq1}(q\rightarrow0)=\mathcal{O}(q^2)$. For $q=0$, $z_{*}=1$ is the only root of $h(z)$. This trivial pole is related to the equilibrium state found for $m=m'=q=0$. For $q\rightarrow0$, the Newton-Raphson sequence of iterated roots will be given by $z_{0}=1$, $z_{1}=z_{2}=\ldots=z_{\infty}=M_{q}+\mathcal{O}(q^{4})$. For any choice of $N\geq1$, it is easy to see that $z_{\infty}(N)$ is a root of the $h_{N}(z)$, thus $z_{*}=\lim_{N\rightarrow\infty}z_{\infty}(N)$ is the leading pole of (\ref{relevantevolution}). Up to fourth order in $q$ this pole can be considered simple because $P\mathcal{E}(z_{*})P=z_{*}+O(q^{4})$. Performing the complex integration of (\ref{relevantevolution}) for $n\gg1$ we obtain the leading PR resonance $\gamma(q)$ \cite{Venegeroles}:
\begin{equation}
\gamma(q)=\ln\sum^{\infty}_{j=0}\Psi_{j}(q)+\mathcal{O}(q^{4})\,.
\label{LPR}
\end{equation}
The relaxation rate (\ref{LPR}) is called slow because $\gamma(q)=\mathcal{O}(q^{2})$ for small wave number $q$.

\section{Faster Relaxation Rate\label{LPres}}

Likewise the leading resonance corresponding to the diffusive modes of the momentum variable $I$ leads the exponential relaxation of distribution functions, leading angular resonances have a important role in the exponential decay of angular correlation functions
\begin{equation}
C_{uv}(n)=\left\langle u|U^{n}|v\right\rangle\sim e^{-n\gamma},
\label{corr}
\end{equation}
where $u$ and $v$ are two any observables at the same instant of time and $n$ is sufficiently large. Let us consider the analysis of the transition elements $Q_{1}U^{n}Q\equiv\left\langle m,0|U^{n}|m',q'\right\rangle$. The expansion of $Q_{1}R(z)Q$ can be written as
\begin{equation}
Q_{1}\frac{1}{z-U}Q=\sum_{i=1}^{\infty}z^{-(i+1)}\phi_{i},
\label{QRQ}
\end{equation}
where $\phi_{i}\equiv Q_{1}U^{i}Q$. The analysis becomes simpler for the LRN case, where $\mathcal{G}_{\lambda}(r=1,x)=\delta_{\lambda,x}$, and for which we have the following first three $\phi_{i}$ coefficients:
\begin{eqnarray}
\phi_{1}&=&\sum_{m'}\mathcal{J}_{m-m'}(-mK),\label{phi1}\\
\phi_{2}&=&\sum_{\lambda}\mathcal{J}_{m-\lambda}(-mK)\sum_{m'}\mathcal{J}_{\lambda-m'}[-(m+\lambda)K],\label{phi2}\\
\phi_{3}&=&\mathcal{J}_{2m}(-mK)\sum_{m'}\mathcal{J}_{-(m+m')}(mK)+\Gamma_{m}(K)+\mathcal{O}(\mathcal{J}^{3})\label{phi3},
\end{eqnarray}
where
\begin{eqnarray}
\Gamma_{m}(K)&\equiv&\mathcal{J}_{m}^{2}(-mK)+\mathcal{J}_{0}(-mK)\mathcal{J}_{m}(-mK)\nonumber\\
&\,&+\sum_{m'}\mathcal{J}_{m-m'}(-mK)\left\{\mathcal{J}_{m+2m'}[-(m+m')K]+\mathcal{J}_{m'}[-(m+m')K]\right\}.
\label{gammaK}
\end{eqnarray}
In the calculation of (\ref{phi1})-(\ref{phi3}), as well as in the calculations that follow, it is crucial to consider the following addition rule 
\begin{equation}
\sum_{m'}\mathcal{J}_{m-m'}(x)=\sum_{\lambda}\mathcal{J}_{m-\lambda}(x)-\mathcal{J}_{m}(x)=1-\mathcal{J}_{m}(x).
\label{eqsum}
\end{equation}
Notice that, including $l=0$, we have $\sum_{l}\exp(-ilt)=2\pi\sum_{l}\delta(t-2\pi l)$. Hence, the identity (\ref{eqsum}) holds due to $\sum_{\lambda}\mathcal{J}_{\lambda}(x)=\exp[ixf(0)]=1$ for $f(0)=0$. Such a result was only known for the particular case of Bessel functions of first kind by means of its generating function \cite{Abramowitz}.

For sufficiently high values of $K$ we expect that the coefficients $\phi_{i}$ become negligible as $i$ increases. Thus, in a first approximation, we can truncate the right hand side of (\ref{QRQ}) at $i=3$ and rewrite it in the following rational form
\begin{equation}
\frac{\phi_{1}z^{2}+\phi_{2}z+\phi_{3}}{z^{4}}\approx\frac{z^{-4}}{\psi_{0}+\psi_{1}z+\psi_{2}z^{2}},
\label{rational}
\end{equation}
whose coefficients $\psi_{i}$ are given in terms of $\phi_{i}$ as
\begin{equation}
\psi_{0}=\frac{1}{\phi_{3}},\qquad\psi_{1}=-\frac{\phi_{2}}{\phi_{3}^{2}},\qquad\psi_{2}=\frac{\phi_{2}^{2}}{\phi_{3}^{3}}-\frac{\phi_{1}}{\phi_{3}^{2}}.
\end{equation}
The right hand side of (\ref{rational}) is, in a first approximation, the analytical extension of the series representation of $Q_{1}R(z)Q$, valid in the chaotic regime. The non null poles of the projected resolvent (\ref{rational}) form the leading resonances of the PF operator $Q_{1}UQ$. First, we have $\phi_{1}=1-\mathcal{J}_{m}(-mK)$ due to (\ref{eqsum}), thus $\phi_{1}\neq0$ unless $K=-1$ for the particular case of the sawtooth map $f(\theta)=\theta$. Considering $\phi_{1}=1+\mathcal{O}(\mathcal{J})$ as the dominant term, $\phi_{2}$ (for $\lambda=m'=-m$) and $\phi_{3}$ must be the $\mathcal{O}(\mathcal{J})$ perturbative terms of the $\phi$-expansion (\ref{rational}). Neglecting only $\mathcal{O}(\mathcal{J}^{3})$ terms on the $\phi_{i}$ coefficients, the poles of the rational form (\ref{rational}) will be given by
\begin{equation}
z_{\pm}=\pm\sqrt{\frac{\phi_{3}}{\phi_{1}}}-\frac{1}{2}\frac{\phi_{2}}{\phi_{1}}+\mathcal{O}(\mathcal{J}^{3/2}).
\label{zpm}
\end{equation}
The ratio $\phi_{2}/\phi_{1}$ can be considered only as $\mathcal{J}_{2m}(-mK)$, and its $\mathcal{O}(\mathcal{J}^{2})$ terms can be neglected. On the other hand, the ratio $\phi_{3}/\phi_{1}\sim\mathcal{J}_{2m}(-mK)$ must be considered up to $\mathcal{O}(\mathcal{J}^{2})$ terms, given $\mathcal{O}(\mathcal{J}^{1/2})$ and $\mathcal{O}(\mathcal{J})$ corrections, respectively. Thus, the leading angular resonance, represented in the exponential form as $|z|=\exp(-\gamma)$, will be
\begin{equation}
\gamma=-\ln\max_{m}\left|\sqrt{\mathcal{J}_{2m}(-mK)+\Gamma_{m}(K)}\pm\frac{1}{2}\mathcal{J}_{2m}(-mK)\right|
\label{lnPR}
\end{equation}
with $\Gamma_{m}(K)$ given by (\ref{gammaK}). Note that, for odd impulse function $f(\theta)$, the leading resonance (\ref{lnPR}) is invariant under the change $m\rightarrow-m$.

\begin{figure}[ht]
\resizebox{0.7\linewidth}{!}{\includegraphics*{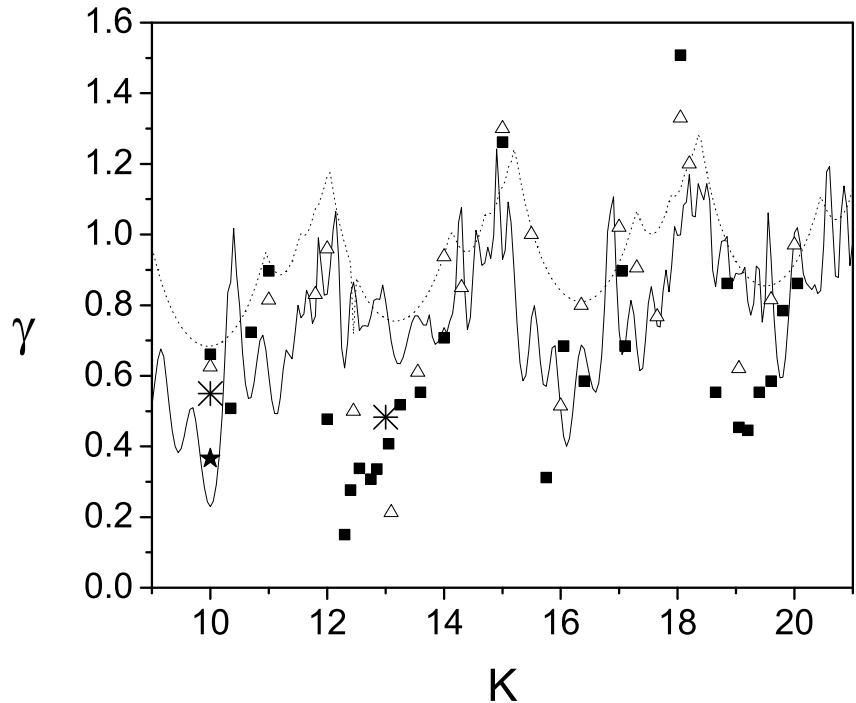}}
\caption{Theoretical leading resonance $\gamma$ (solid line) calculated for the standard map compared with its asymptotic value $\gamma_{\infty}$ (dotted line) and several numerical calculations. Here, $(\blacksquare)$ and $(\triangle)$ represent the resonances calculated from $C_{1,1}$ and $C_{1,2}$ correlations, respectively, by Khodas et al. \cite{Khodas}, $(\divideontimes)$ represents the resonances calculate by Blum and Agam \cite{Blum}, and $(\bigstar)$ is the intermediary value calculated by Florido et al. \cite{Florido}.}
\label{smangresg}
\end{figure}

For very large values of $K$, the leading resonance (\ref{lnPR}) tends to the following value:
\begin{equation}
\gamma_{\infty}=-\ln\max_{m}|\sqrt{\mathcal{J}_{2m}(-mK)}|,
\label{lnPR0}
\end{equation}
obtained in a different way by Khodas et al. for the standard map particular case \cite{Khodas}. It is important to check the limits of validity of each approximation and its respective adequacy to the numerical values existing in the literature. In Fig. \ref{smangresg} we compare, for the standard map, the resonance (\ref{lnPR}) with its asymptotic value (\ref{lnPR0}). For sufficiently large times, Khodas et al. \cite{Khodas} calculate numerically the correlation (\ref{corr}) for $u$ and $v$ proportional to $\exp(im\theta)$ and for some combinations of modes $\left\{m,m'\right\}$, where $C_{uv}\equiv C_{m,m'}$ in this choice. Once the resonance (\ref{lnPR}) is dominated by initial mode $m=1$, we select $C_{1,1}$ and $C_{1,2}$ as the best simulated correlations. However, these numerical values have only a qualitative character for sake of comparison, since resonance (\ref{lnPR}) leads the decay of correlations only for very large times, when the numerical signal is too weak \cite{Khodas}. Moreover, the precise composition of the observable $v$ as a possible superposition of modes $\left\{m'_{i}\right\}$ for which $C_{1,v}$ decay through (\ref{lnPR}) is not known. On the other hand, we also include the two values of leading resonances calculated numerically by diagonalization of $U$ for $K=10$ and $K=13$ by Blum and Agam \cite{Blum}, in addition to the leading intermediary value calculated by Florido et al. for $K=10$ (assumed here between $z=0.672$ and $z=0.715$) \cite{Florido}. By comparing all these results, the theoretical result (\ref{lnPR}) gives a better qualitative fit with the numerical values even for high values of $K$ and, besides, it reveals a more intrincated structure of peaks for the maximal resonance.

The sawtooth map $f(\theta)=\theta$ is the only particular LRN case for which the perturbative scheme presented above breaks down \cite{Sano2}. This occurs in such case due to $\mathcal{J}_{m}(x)=\frac{\sin[\pi(m-x)]}{\pi(m-x)}=1$ for $m=x$ and interger values of $K$. For example, besides $\phi_{1}=1$ for all integer $K\neq-1$, $\phi_{2}$ can be rewritten as
\begin{equation}
\phi_{2}=1-\sum_{\lambda}\mathcal{J}_{m-\lambda}(-mK)\mathcal{J}_{\lambda}[-(m+\lambda)K].
\label{p2n}
\end{equation}
For integer $K$, the sum in the right hand side of (\ref{p2n}) vanishes unless $-K=\frac{\lambda}{m+\lambda}=\frac{m-\lambda}{m}$, which gives $|\lambda/m|=g_{*}$ or $|\lambda/m|=g_{*}^{-1}$, where $g_{*}=(\sqrt{5}-1)/2$ is the golden mean. Hence, we also have $\phi_{2}=1$ for all integer $K$. This suggests that $\phi_{i}=1$ for almost all integer $K$. If this hypothesis is true, we then have as analytical continuation of the resolvent for $|z|>1$
\begin{equation}
Q_{1}\frac{1}{z-U}Q=\frac{1}{z^{2}}\sum_{j=0}^{\infty}\frac{1}{z^{j}}=\frac{1}{z(z-1)},
\end{equation}
according to (\ref{QRQ}). Thus, $z=1$ corresponds to the invariant density and all the others resonances are infinitely degenerated at $z=0$. This particular result was demonstrated by Sano for all positive integers $K$ by using the Fredholm determinat of $U$ \cite{Sano}.

\section{Concluding Remarks\label{Concluding}}

In conclusion, we have presented a method to determine analytically leading Pollicott-Ruelle resonances which is applicable to a general class of area-preserving maps, including mixed systems. Such resonances are obtained through the resolvent of the PF operator by using projection operator techniques. In particular, we calculate the leading resonance related to the slow modes of relaxation, which corresponds to the diffusive process, as well as the leading resonance related to the faster modes of relaxation. In this last case, our perturbative analysis was performed only for systems with linear rotation numbers, although it can be similarly applied for nonlinear ones.

The analytical results obtained here have been compared with theoretical and numerical calculations existing in the literature. The resonance (\ref{lnPR}) was calculated in a systematic way in which correction terms of order $\mathcal{O}(\mathcal{J})$ produce a more intrincated structure of peaks for the standard map case even for high values of $K$, as can be seen in Fig. \ref{smangresg}. Despite the absence of estimates of errors in the numerical results, the agreement with the theoretical result (\ref{lnPR}) is reasonable. We have also investigated particular characteristics of the sawtooth map that are incompatible with the perturbative approach developed in the Sec. \ref{LPres}. Our analysis points towards the accordance between our hypotesis and the results presented in \cite{Sano}.

\subsection*{Acknowledgments}

The author thanks M.M. Sano for kindly sharing his numerical results, and A. Saa, E. Abdalla, W.F. Wreszinski, R. da Rocha, and E. Gu\'eron for helpful discussions. This work was supported by UFABC.


\begin{thebibliography}{9}


\bibitem{Gaspard} P. Gaspard, {\it Chaos, Scattering and Statistical Mechanics} (Cambridge University Press, Cambridge, England, 1998).

\bibitem{CvitanovicBook} P. Cvitanovi\'c, R. Artuso, R. Mainieri, G. Tanner, and G. Vattay, {\it Classical and Quantum Chaos} (Niels Bohr Institute, Copenhagen, 2004, www.nbi.dk/ChaosBook/).

\bibitem{Pollicott} M. Pollicott, Invent. Math. {\bf 81}, 413 (1985); M. Pollicott, Invent. Math. {\bf 85}, 147 (1986).

\bibitem{Ruelle} D. Ruelle, Phys. Rev. Lett. {\bf 56}, 405 (1986); D. Ruelle, J. Stat. Phys. {\bf 44}, 281 (1986).

\bibitem{Hasegawa2} H.H. Hasegawa and W.C. Saphir, Phys. Rev. A {\bf 46}, 7401 (1992).

\bibitem{Khodas} M. Khodas and S. Fishman, Phys. Rev. Lett. {\bf 84}, 2837 (2000); M. Khodas, S. Fishman and O. Agam, Phys. Rev. E {\bf 62}, 4769 (2000).

\bibitem{Venegeroles} R. Venegeroles, Phys. Rev. Lett. {\bf 99}, 014101 (2007).

\bibitem{quantum} B.V. Fine, Phys. Rev. Lett. {\bf 94}, 247601 (2005); I. Garcia-Mata, M. Saraceno, and M.E. Spina, Phys. Rev. Lett. {\bf 91}, 064101 (2003); C. Manderfeld, J. Phys. A: Math. Gen. {\bf 36}, 6379 (2003); K. Pance, W. Lu, S. Sridhar, Phys. Rev. Lett. {\bf 85}, 2737 (2000); A.V. Andreev, O. Agam, B.D. Simons, and B.L. Altshuler, Phys. Rev. Lett {\bf 76}, 3947 (1996).

\bibitem{Blum} G. Blum, O. Agam, Phys. Rev. E {\bf 62}, 1977 (2000).

\bibitem{Florido} R. Florido, J.M. Martin-Gonzalez, and J.M. GomezLlorente, Phys. Rev. E {\bf 66}, 046208 (2002).

\bibitem{Weber} J. Weber, F. Haake, P. Seba, Phys. Rev. Lett. {\bf 85}, 3620 (2000);\\ J. Weber, F. Haake, P.A. Braun, C. Manderfeld, and Seba, J. Phys. A: Math. Gen. {\bf 34}, 7195 (2001).

\bibitem{Sano} M.M. Sano, Phys. Rev. E {\bf 66}, 046211 (2002).

\bibitem{Lebouefzeta} P. Leboeuf, Phys. Rev. E {\bf 69}, 026204 (2004).

\bibitem{Lichtenberg} A.J. Lichtenberg and M.A. Lieberman, {\it Regular and Chaotic Dynamics} (Springer, New York, 1992).

\bibitem{Chirikov} B.V. Chirikov, Phys. Reports {\bf 52}, 265 (1979).

\bibitem{nonG} R. Venegeroles and A. Saa, J. Stat. Mech. P01005 (2008).

\bibitem{Bandtlow} O.F. Bandtlow and P.V. Coveney, J. Phys. A: Math. Gen. {\bf 27}, 7939 (1994).

\bibitem {Balescu1} R. Balescu, {\it Statistical Dynamics, Matter out of Equilibrium} (Imperial College Press, London, 1997).

\bibitem{Abramowitz} M. Abramowitz and I.A. Stegun, {\it Handbook of Mathematical Functions} (Dover, New York, 1972).

\bibitem{Sano2} I realized this fact after receiving the numerical results of leading resonances for the sawtooth map calculated by M.M. Sano (private communication).


\bibliographystyle{apsrev}


\end{thebibliography}
\end{document}